\documentclass[peerreview]{IEEEtran}
\usepackage{cite} 
\usepackage{url} 
\usepackage[utf8]{inputenc} 
\usepackage{booktabs} 
\usepackage{graphicx}

\begin{document}
\title{LLM-Assisted Detection and Repair of Hardware Security Vulnerabilities in Verilog Designs}

\author{
    Ethen Santana\\
    Department of Computer Science\\
    Iowa State University\\
    Ames, IA, USA\\
    esantana@iastate.edu
    \and
    John Doe\\
    Department of Electrical Engineering\\
    University of Iowa\\
    Iowa City, IA, USA\\
    jdoe@uiowa.edu
}

\IEEEpeerreviewmaketitle

\begin{center}
{\large
Ethen Santana \quad Gabriel Gyaase \quad Hao Zheng
}

\vspace{0.5em}

Bellini College of Artificial Intelligence, Cybersecurity and Computing\\
University of South Florida\\
Tampa, Fl, USA
\end{center}

\vspace{1em}

\begin{abstract}
Hardware designs, like software, are susceptible to bugs that can introduce security vulnerabilities and create opportunities for malicious exploitation. Unlike software vulnerabilities, however, hardware flaws become permanently embedded in silicon after fabrication, making them difficult or impossible to patch. Many of these weaknesses are categorized under the Common Weakness Enumeration (CWE) framework and include improper access control, exposure of sensitive information, and unintended privilege escalation. To improve the detection of such vulnerabilities, we propose a methodology that leverages a Large Language Model (LLM) to identify potential hardware CWEs directly from hardware designs in Verilog. The proposed approach is evaluated iteratively on a dataset of single-module Verilog designs to assess its effectiveness in detecting hardware security weaknesses. Our results demonstrate the potential of LLMs to augment traditional hardware security analysis by providing automated, scalable assistance for identifying security vulnerabilities during the hardware design process. We open-source our dataset, results, and prompts at \url{https://github.com/JazzCat4/LLM-CWE-Detection}

\end{abstract}

\section{Introduction}
Due to the constant improvement and evolution of software security, computers have become increasingly difficult for adversaries to exfiltrate personal information from. However, even if the software on a device is secure, its hardware can still be prone to vulnerabilities that nullify software-protections. Most circuits can be represented by a hardware description language (HDL), like a programming language. An example of an HDL is Verilog, which can describe circuits at the Register Transfer Level (RTL), which is a combination of the dataflow between circuit components and the intentional functionality of the system (Meng et al. 2025). However, if a circuit with a bug is produced, that bug becomes permanent as the physical circuit is created (Ahmad et al. 2024). As a result, MITRE has created a list called Common Weakness Enumerations (CWE), consisting of different groups of these vulnerabilities on a device's software or hardware based on its design. Hardware-based CWEs can lead to unauthorized data exposure, privilege escalation, and other issues if not properly checked.

According to Meng et al (2025), the main bugs within Verilog RTL involve assignment statements, variable declaration statements, and if-statements, all of which are common within complex circuits. An issue that arises with the increasing number of bugs is that manual debugging has become more tedious. Therefore, hardware designers have started to employ language learning models (LLMs) such as MAGE and AIVRIL as a means of RTL code repair, particularly with locating errors and creating testbenches (Paria and Bhunia, 2026). However, LLMs are prone to hallucination, and lack the sufficient knowledge required to provide repair without needing a human to review it properly for any additional bugs (Alsaqer et al. 2024). Therefore, we propose a new framework to not only make the task of debugging RTL Verilog code less tedious but also provide a guide to automate LLMs to debug and repair RTL code more efficiently.

\section{Motivation}
We created our methodology to simplify the task of locating and fixing bugs for both individuals and LLMs; Finding them earlier throughout the hardware design process makes it easier to fix, and reduces potential costs in replacing faulty hardware in the future. In addition, LLMs have become increasingly relevant in the field of cybersecurity, and hardware designers have begun using LLMs to simplify tasks such as testbench creation. In some cases, LLMs have shown themselves to contribute to more accurate and effective RTL repairs (Alsaqer et al. 2024). However, the main struggle of LLMs in hardware design is that due to their lack of knowledge in the subject and inherent bias against hardware design, they often struggle with more complex circuits. (Paria and Bhunia, 2026). This issue is contributed due to under-constrained repair: A bug in RTL code allows the LLM to determine multiple different methods on how to fix it, regardless of if they are relevant to the design (Mastora and Sullivan, 2026). Underconstrained repair is primarily caused by three main factors: prompting, data scarcity, and long context processing, all of which can interfere with the learning of the LLM, preventing it from obtaining a true answer (Liu et al, 2026). Therefore, this framework is also meant to ensure that LLMs obtain proper fine-tuning to ensure that they will be able to assess hardware design more accurately, and eventually reduce the requirment of human assistance.

\section{Related Works}
The Verilog LLM-Aided Vulnerability Detection framework (VerilogLAVD) by Long et al. (2026) explores the idea of LLMs detecting vulnerabilities by creating a property graph based on its structure and detecting its vulnerabilities based on its time-related edges. The researchers assess the performance of multiple LLMs using this method through their quality, reliability and practicality to determine that this method has greatly improved LLM’s ability to detect vulnerabilities within Verilog code. However, one of the limitations within their research was the capabilities of the LLM they were using, as it struggled with tracing back vulnerabilities for longer periods of time. According to a survey by Liu et al. (2026), one of the three main challenges of LLMs in hardware is long context processing, as if it surpasses its intended capacity, the LLM will lose information, forgetting its original “thought process”. We solved this issue by using the prompts to reference previous parts of the conversation, which include important details such as RTL design rules. Although this will not be the same line of thought, reviewing and considering its previous outputs will lead it to create a more in-line solution for the next steps of the methodology.

Qi et al’s Verilog Retrieval Automation Generation (VeriRAG) focuses on using retrieval augmented generation to help LLMs improve RTL design so that they can become easier to test for vulnerabilities (2026). VeriRAG achieves this by comparing a given RTL design with the most similar RTL from an example dataset, then using that and a given guide by the researchers to detect manual errors in the RTL design and create fixes for them. A shared aspect between our methodology and VeriRAG is that further knowledge on the general subject of RTL is incrementally added over time, but the key difference is that VeriRAG gives the LLM this knowledge to apply to the RTL module, while our methodology seeks to have the LLM approach that answer by itself.

Our methodology hopes to build on the ideas of VeriRAG and VerilogLAVD by analyzing hardware CWEs to allow both people and LLMs to better understand and reiteratively improve on their Verilog RTL code to minimize the number of possible exploits within circuits.

\section{Methodology} \label{sec:criteria}
Our methodology was designed through the analysis of CWE entries (MITRE, 2025), previous works, and iterative testing with the goal of providing an effective and efficient method to identify vulnerabilities that those with limited knowledge can utilize. We have limited our methodology and research to the listed CWEs on MITRE's 2025 Most Important Hardware Weaknesses list to target more commonly seen vulnerabilities. Two CWE entries on this list (CWE-1421 and CWE-1423) were omitted due to these vulnerabilities arising from microarchitectural design considerations.\\

\textbf{Complete list of CWEs:}
\begin{itemize}
    \item CWE-226: Sensitive Information in Resource Not Removed Before Reuse
    \item CWE-1189: Improper Isolation of Shared Resources on System-on-a-Chip (SoC)
    \item CWE-1191: On-Chip Debug and Test Interface With Improper Access Control
    \item CWE-1234: Hardware Internal or Debug Modes Allow Override of Locks
    \item CWE-1247: Improper Protection Against Voltage and Clock Glitches
    \item CWE-1256: Improper Restriction of Software Interfaces to Hardware Features
    \item CWE-1260: Improper Handling of Overlap Between Protected Memory Ranges
    \item CWE-1262: Improper Access Control for Register Interface
    \item CWE-1300: Improper Protection of Physical Side Channels
\end{itemize}

As Figure 1 describes, when the LLM receives an RTL design, the role and features of the module are first identified. The module is then analyzed to determine important assets, behavior, data, and control flow. A CWE-driven review is then performed utilizing the results of the analysis and module to identify potential vulnerabilities. A testbench is then created targeting potential and suspected vulnerabilities. The module is simulated using the testbench, and repaired utilizing CWE-Design rules until the module passes simulation. We utilized Microsoft Copilot as the LLM to judge performance; Other models may yield different results.

\begin{figure}
    \centering
    \includegraphics[width=1\linewidth]{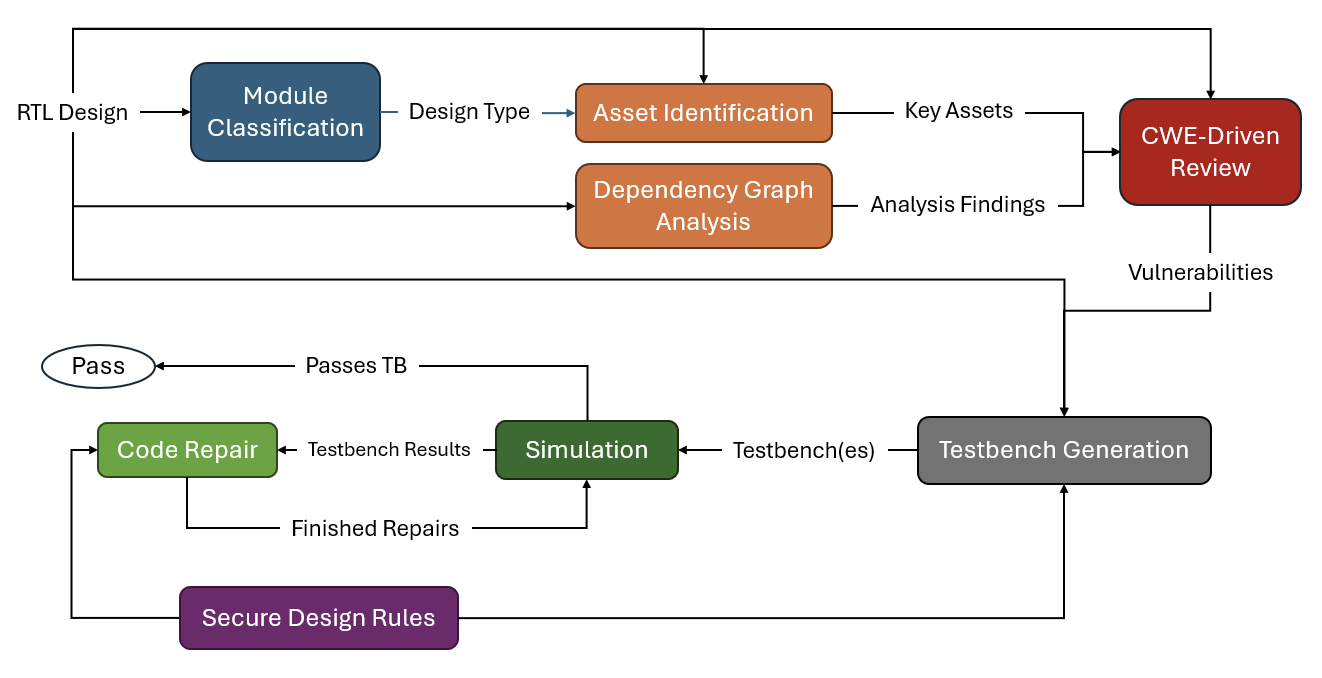}
    \caption{High-Level Graph of Methodology}
    \label{fig:placeholder}
\end{figure}

\begin{figure}
    \centering
    \includegraphics[width=1\linewidth]{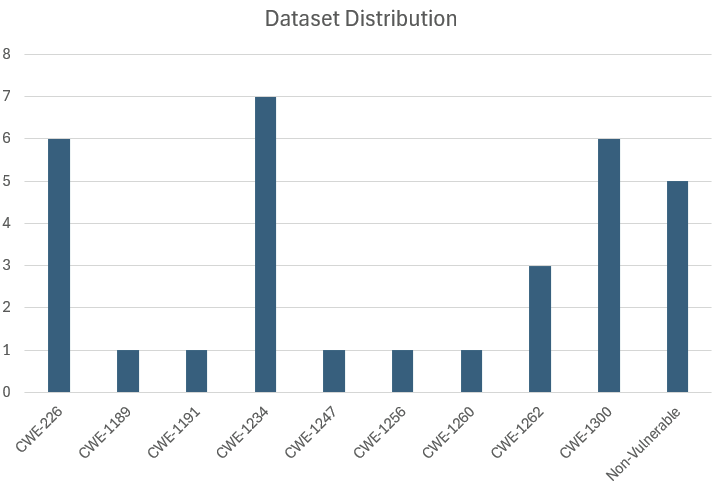}
    \caption{Distribution of CWEs for dataset}
    \label{fig:placeholder}
\end{figure}

Our methodology was tested on a dataset of thirty-two Verilog single-module designs. Of these modules, twenty-seven contained a corresponding CWE in its design. The remaining five modules were used to test the validity of our methodology and the AI model's ability to determine if a module contained vulnerabilities on modules that had no CWE. The Verilog modules were designed through study of the common causes of corresponding CWEs. Some modules were also obtained from the dataset of a previous study (Qi et al., 2026). The distribution of the modules are as follows shown in Figure 2.\\

\subsection{Module Classification}
The role of the module and any potential features are first identified. The AI model is given the potentially buggy module, along with a list of Module Types, Features, and their corresponding potential CWEs (Figure 3). The AI model is instructured to identify any of the features on the list present in the given RTL design. It is possible to have multiple features in the design depending on the complexity of the module.

By identifying these features, we are able to narrow down the amount of CWEs to search for when conducting the CWE-Driven review; For example, it would be unnecessary to check for improper JTAG authentication in a module that does not have a JTAG interface. Reviewing all CWEs also increases the risk of providing an overwhelming amount of information to the AI, potentially degrading its performance. The selected CWEs are recorded as potential vulnerabilities that will be reviewed in the CWE-Driven review.

\begin{figure}
    \centering
    \includegraphics[width=1\linewidth]{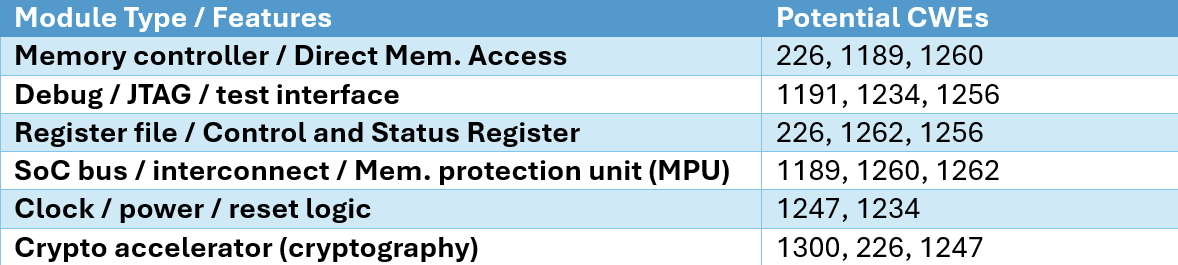}
    \caption{List of Module Types, Features and Corresponding CWEs}
    \label{fig:placeholder}
\end{figure}

\subsection{Asset Identification}
Next, any important assets are to be identified. The AI model is given the RTL design, the previously found module type, and features, and is instructed to identify any important assets, behavior, and functionality of the module. This includes the important assets such as keys, boundaries between privilege domains, clock behavior, reset schemes, adversary capabilities, security rules, and Finite State Machine states.

Identifying the important assets of a module will help guide the AI model into the kinds of problems the module may have, along with verifying its understanding of the module's behavior itself. This also aids in the CWE-Driven review, as the AI has a better understanding of where to look for potential vulnerabilities in the RTL design, improving its efficiency and effectiveness.

\subsection{Dependency Graph Analysis}
A Program Dependency Graph (PDG) is then created and analyzed. Given the RTL design, the AI model is instructed to generate a PDG that models the data and control flow of the design. The nodes of the graph consists of any signals, registers, and gates. The edges of the graph consist of data-flow and control-dependency relationships. Based off of the generated graph, the AI model is then instructed to perform the following analyses:\\
\begin{itemize}
    \item \textbf{Reachability Analysis: }Determine whether one node can be reached from another through a path in the graph.
    \item \textbf{Dominance Analysis: }Determine whether one node must always be encountered before another node.
    \item \textbf{Taint Propagation Analysis: }Track how information originating from an untrusted source flows through the design.
\end{itemize}

The AI model returns the results from the performed analyses, consisting of the behaviors, controls, and potential vulnerabilities of the design. Through graph creation and analysis, the AI model is able to better define the behavior and design of the module, and potentially find vulnerabilities prior to the CWE-Driven Review. These results are all fed into the CWE-Driven review to aid in vulnerability identification in the module.

\subsection{CWE-Driven Manual Review}
Utilizing the module and the results from the analysis, a CWE-driven review is performed. The AI model is given the RTL design, the identified key assets, the results from the PDG analysis, and relevant CWE guides. The AI model is instructed to utilize the given inputs and follow all provided CWE guides to identify vulnerabilities in the module.

CWE guides are chosen based on suspected or potential CWEs determined from the results of the module classification, asset and behavior identification, and graph analysis. The CWE guides were developed through analysis of their respective CWE entries with regard to their common causes and detection methods, with the goal of aiding in the detection of the respective CWE. The LLM is given only relevant CWE guides based off of its previous findings as to not overwhelm the model with too much information, potentially degrading its performance. Each CWE guide contains a brief description of the vulnerability, questions to guide thinking, common causes of the vulnerability, and a checklist to aid in identification. The AI models returns a list of found vulnerabilities with its corresponding CWE entry, along with a description of the issue in the module. These results are then used to generate a testbench to verify these vulnerabilities.

The complete list of CWE guides can be found in the prompting directory of the repository.

\subsection{Testbench Generation}

Utilizing the previously found vulnerabilities, a testbench for the RTL design is created. The AI model is given the RTL design, the found vulnerabilities (if any), and a list of CWE secure design rules, and is instructed to generate a testbench that targets all found vulnerabilities and the validity of the given secure design rules for each CWE. Similar to the CWE guides, a list of CWE design rules were created for each CWE. These design rules were developed through analysis of their corresponding CWE entries, including their mitigation methods, and common mistakes that lead to these vulnerabilities. All testbench tests should be reviewed for potential errors, as a false pass / fail may negatively impact performance and code repair.

The complete list of CWE design rules can be found in the prompting directory of the repository.

\subsection{Simulation}
The generated testbench is then compiled and ran to test the given RTL design under the generated tests. If all tests pass, the given RTL module passes inspection, and all testing and analysis is completed. If one or more tests fail, the RTL design will then undergo a code repair.

\subsection{Code Repair}
Code repair is performed utilizing the failed test cases resulted from simulation. The AI model is given the RTL design, all failed testbench tests, and the list of CWE design rules. The AI model is instructed to fix the module based off of the failed tests from the testbench simulation, while utilzing the CWE design rules as a guide for code repair. The AI model is also instructed to generate an updated testbench if changes to the RTL design were made that make it incompatible to the previously generated testbench (e.g. a new input). Once the fixed module (and potentially updated testbench) is generated, the simulation step is repeated testing the repaired RTL design. Code repair was performed a maximum of three times; If the repaired module continued to fail the testbench after the third iteration, the test was marked as a failure.

\section{Results}
Analysis reveals that the AI model demonstrated strong capabilities in understanding and reasoning about Verilog hardware designs and CWEs, but its performance varied depending on the task being performed.

The module classification produced favorable results. Across the evaluation, the AI correctly identified the functional type of the Verilog module in nearly all cases, with only one case of misclassifications observed.

The AI consistently performed well during the early stages of analysis. During asset identification, it was generally able to identify the primary security-relevant assets within a module and explain their intended functionality and importance. In many cases, the model correctly recognized the hardware vulnerability before performing the CWE-driven review. Likewise, the AI accurately identified the behavior of modules and provided meaningful explanations of Project Dependency Graph (PDG) relationships and results from its analysis, showcasing its ability to reason about both the structural and behavioral aspects of the design.

The greatest weakness was observed during testbench generation. AI-generated testbenches frequently failed to validate the intended security properties of the design. Common issues included incorrect use of the Device Under Test (DUT), incomplete or ineffective test cases, and verification procedures that did not meaningfully determine whether the vulnerability was present or had been mitigated. In several instances, the generated testbenches also failed to properly reset the DUT between each test case, allowing the states from the previous tests to influence subsequent results, further reducing the reliability of the testbench.

Code repair produced mixed results. While the AI was often capable of generating syntactically correct patches that addressed the identified vulnerability, repairs occasionally modified the intended functionality of the original design. The reasoning from this stems from the lack of context given to the AI; The AI is given the module without any information about its intended behavior. Rather than implementing the minimal changes necessary to mitigate the weakness, the model sometimes introduced additional security mechanisms that were not required by the specification. For example, the AI occasionally added privilege-level inputs or access-control logic to modules whose intended functionality did not require privilege enforcement. These modifications improved security in a general sense but altered the original design requirements, highlighting the tendency of the model to favor security enhancements over preserving the original hardware behavior. In total, twenty-seven of the thirty-two total tests conducted resulted in a pass; an 84\% success rate.
\subsection{Non-Vulnerabilities}
Of the thirty-two total modules, five Verilog modules did not contain vulnerabilities, and were tested to assess the AI's ability to distinguish secure designs from vulnerable ones. In these cases, the model showcassed a high false-positive rate. Across all five non-vulnerable modules tested, the AI identified at least one vulnerability, despite the modules being intentionally designed without security flaws.

Many of the reported vulnerabilities stemmed from the AI recommending security features that were unnecessary for the intended functionality of the module. For example, the model frequently suggested the addition of lock bits, write-once protections, or privilege-based access controls even when the design did not require these features. The AI often treated the absence of additional security mechanisms as a vulnerability.

The AI also showcased difficulty interpreting intended module behavior during testbench generation. In several cases, generated testbenches labeled intended functionality as incorrect, such as flagging the absence of data scrubbing after returning to a particular state even though this behavior was necessary for the module's opeartion. These incorrect assumptions resulted in failed test cases that the AI mistook for genuine security weaknesses. Because the failures originated from incorrect tests, code repair was not performed, as doing so would likely have altered the functionality of otherwise correct designs.

Consistent with the observations made for vulnerable modules, the quality of AI-generated testbenches remained a significant issue. Testbenches frequently contained ineffective or inappropriate security tests that did not accurately verify vulnerabilities, resulting in misleading testbench outcomes. Of the five non-vulnerable modules evaluated, three produced testbenches resulting in all tests passing. The remaining two testbenches were poorly designed and did not contain well-designed tests, which resulted in failed tests, further demonstrating that inaccurate verification was a major contributor to the AI's false-positive assessments.
\section{Failed Tests}
\begin{itemize}
\item\textbf{226\_4}\\
CWE-226 was correctly identified in this design. However, during simulation, the generated testbenches consistently produced uninitialized output values (x). Across three iterations of code repair and testbench regeneration, this issue persisted and prevented the testbenches from producing meaningful verification results. Consequently, the repaired module was unable to pass the generated tests.

\item\textbf{1300\_5}\\
The AI model failed to classify the module as a Crypto Accelerator. Instead, the module was classified as Clock/Power/Reset Logic and Register File/CSR. Although these classifications were not entirely incorrect, they failed to capture the module's primary functionality. Consequently, CWE-1300 was omitted from the CWE-Driven Review. The subsequent code repair improved aspects of the module's security; however, the underlying CWE-1300 vulnerability remained unresolved.

\item\textbf{1300\_6}\\
CWE-1300 was correctly identified in this design. However, the generated testbench failed to provide meaningful information regarding the module's potential vulnerabilities. During code repair, an error state was added to the design to activate when a glitch or reset attack was detected. This error state did not activate as intended, resulting in failed tests. Two additional iterations of code repair were performed, but the issue persisted. Ultimately, the repaired module was unable to pass the generated testbench.

\item\textbf{Non-Vulnerability\_2}\\
Although this module contained no known vulnerabilities, the AI identified several potential security issues, including inadequate data scrubbing, the absence of error or safe states, and unprotected internal registers. The generated testbench included tests that incorrectly flagged intended module functionality as vulnerable, demonstrating a misinterpretation of the module's intended behavior. Although the simulation failed, code repair was not performed because the identified issues did not represent actual vulnerabilities and modifying the design would likely have altered its intended functionality.

\item\textbf{Non-Vulnerability\_3}\\
Similar to Non-Vulnerability\_2, the AI identified several vulnerabilities in this module despite the design containing no known security weaknesses. These included unscrubbed reuse of output variables and the absence of lock bits. The generated testbench attempted to verify the reported vulnerabilities; however, many of the tests were inaccurate. Several tests incorrectly flagged intended module behavior, while others modeled an attacker as having direct control over protected inputs, vulnerabilities beyond the scope of the RTL design. No code repair was performed because the identified issues did not represent actual vulnerabilities and the generated tests did not provide sufficient evidence to justify modifying the module.
\end{itemize}

\section{Conclusions and Recommendations}
The results of this study demonstrate that the proposed methodology can support LLM-based analysis of RTL designs and hardware vulnerabilities. The LLM showed strong capabilities in interpreting module behavior, identifying security-relevant assets, and recognizing known CWEs. However, the failed tests revealed several limitations, particularly in RTL reasoning, testbench generation, and vulnerability verification. These errors suggest that factors such as limited RTL-specific knowledge and hallucination may negatively affect the reliability of LLM-based hardware security analysis.

In Tests 226\_4 and 1300\_6, the LLM correctly identified the targeted vulnerabilities but failed to generate effective RTL repairs and testbenches, even after multiple iterations. These results indicate limitations in the model's ability to translate vulnerability knowledge into functional RTL implementations and meaningful verification procedures. In Test 1300\_5, the LLM failed to identify the module's primary functionality, resulting in CWE-1300 being omitted from the CWE-Driven Review. Instead, the model focused on other perceived security concerns and generated repairs that did not address the intended vulnerability. This behavior may be attributed to model bias or hallucination caused by under-constrained repair prompts, which can lead the model to identify and address unintended issues (Mastora and Sullivan, 2026). Similar behavior was observed in Non-Vulnerability\_2 and Non-Vulnerability\_3, where the LLM reported vulnerabilities in modules that contained no known security weaknesses and misinterpreted intended module behavior as insecure.

Future work should evaluate the methodology across a larger and more diverse set of CWEs and RTL designs to better assess its generalizability and accuracy. Prompting strategies should also be refined to reduce hallucinations and constrain repairs to the identified vulnerability and intended functionality of the module. Additionally, providing the LLM with explicit information regarding the module's purpose and expected behavior may improve module classification, vulnerability analysis, and RTL repair. Overall, these findings demonstrate both the potential and current limitations of LLMs for RTL security analysis. The proposed methodology may help guide the development and fine-tuning of LLM-based hardware security tools, improving their ability to identify, verify, and mitigate CWEs while preserving the intended behavior of RTL designs.

\end{document}